\documentclass[aps,prl,superscriptaddress,twocolumn]{revtex4-1}
\usepackage{graphicx}
\usepackage{amsmath}
\usepackage{amsfonts}
\usepackage{bm}
\usepackage{color}

\begin{document}

\title{Multiferroic Quantum Criticality}

\author{Awadhesh Narayan}
\affiliation{Materials Theory, ETH Zurich, Wolfgang-Pauli-Strasse 27, CH 8093 Zurich, Switzerland}

\author{Andr\'{e}s Cano}
\affiliation{Materials Theory, ETH Zurich, Wolfgang-Pauli-Strasse 27, CH 8093 Zurich, Switzerland}
\affiliation{Institut N\'eel, CNRS \& Univ. Grenoble Alpes, 38042 Grenoble, France}

\author{Alexander V. Balatsky}
\affiliation{NORDITA, Roslagstullsbacken 23, SE-106 91 Stockholm, Sweden}
\affiliation{Institute for Materials Science, Los Alamos, NM 87545, USA}
\affiliation{Department of Physics, University of Connecticut, Storrs, CT 06269, USA}

\author{Nicola A. Spaldin}
\affiliation{Materials Theory, ETH Zurich, Wolfgang-Pauli-Strasse 27, CH 8093 Zurich, Switzerland}

\date{\today}

\begin{abstract}
The zero-temperature limit of a continuous phase transition is marked by a quantum critical point, which can generate exotic physics that extends to elevated temperatures~\cite{sachdev2011}. Magnetic quantum criticality is now well known, and has been explored in systems ranging from heavy fermion metals~\cite{gegenwart2008quantum} to quantum Ising materials~\cite{coldea2010quantum}. Ferroelectric quantum critical behaviour has also been recently established~\cite{rowley2014ferroelectric}, motivating a flurry of research investigating its consequences~\cite{edge2015quantum,stucky2016isotope,rischau2017ferroelectric,chandra2017prospects}. Here, we introduce the concept of multiferroic quantum criticality, in which both magnetic and ferroelectric quantum criticality occur in the same system. We develop the phenomenology of multiferroic quantum critical behaviour, describe the associated experimental signatures, and propose material systems and schemes to realize it.
\end{abstract}

\maketitle

\begin{figure*}
\includegraphics[scale=0.6]{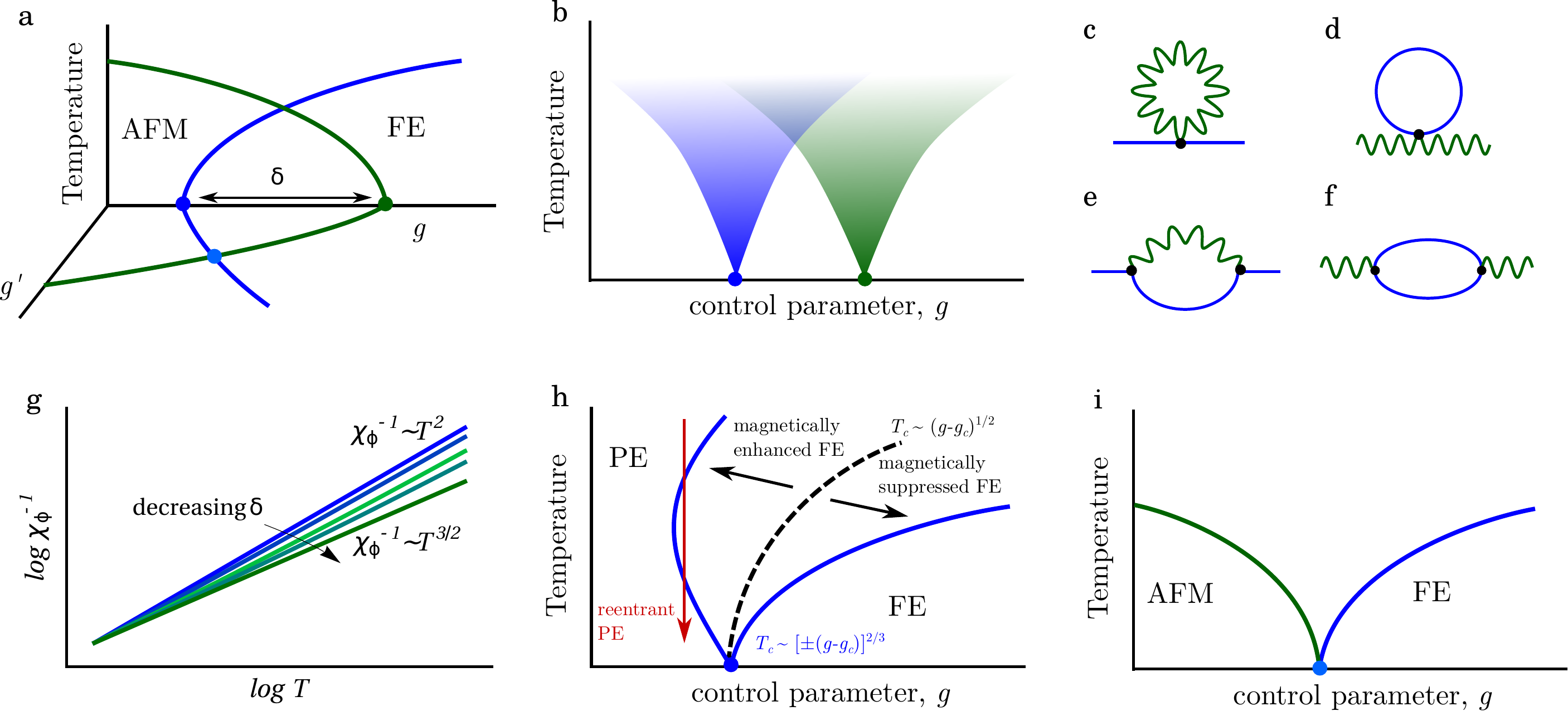}
  \caption{\textbf{Phenomenology and identification of multiferroic quantum criticality.} (a) Magnetic and ferroelectric quantum critical points generically occur at different values of a control parameter, $g$. The separation between the two critical points, $\delta$, can be tuned by an additional external variable, $g'$, and can lead to an ``accidental'' bicritical point. (b) A schematic of quantum criticality regions arising from the zero temperature magnetic and ferroelectric critical points. Note the overlapping region in which magnetoelectric quantum critical excitations are expected. Feynman diagrams for one loop correction to the susceptibilities with (c)-(d) biquadratic coupling and (e)-(f) dynamical coupling. The solid and wavy lines represent the propagators for ferroelectric and magnetic order parameters, respectively. (g) Temperature scaling of inverse dielectric susceptibility, $\chi_{\phi}^{-1}$, for biquadratic coupling with decreasing separation between the quantum critical points, $\delta$. A crossover from an exponent of 2 to 3/2 is predicted as $\delta$ decreases. (h) Modification of the ferroelectric phase diagram due to the MFQC that accompanies this crossover. The ferroelectric phase can be either enhanced or suppressed by the quantum-critical magnetic degrees of freedom, which can lead to a reentrant behavior of the paraelectric (PE) phase as a function of the temperature (red arrow). In both cases, the critical exponents of the magnetic sector will take over the scaling of the ferroelectric $T_c$ with the control parameter.(i) A coupled bicritical point can occur when the two quantum critical points are coincident with the variation of a single control parameter.}  \label{schematic}
\end{figure*}

In conventional thermal phase transitions, ordered phases of matter transition to disordered phases with increasing temperature. Examples include the ferromagnet to paramagnet, or conventional superconductor to normal metal transitions. In a classical universe, zero-temperature fluctuation-driven phase transitions can not occur because of the absence of thermal fluctuations. Quantum mechanics offers richer possibilities, since the quantum fluctuations that occur even at zero temperature can give rise to zero-temperature phase transitions on varying a non-thermal control parameter, such as pressure or doping [Fig.~\ref{schematic}(a)]; the critical point in such a quantum phase transition is called a quantum critical point (QCP)~\cite{sachdev2011}. 

Although the QCP that separates the ordered and disordered phases occurs by definition only at zero temperature, it strongly influences finite-temperature behavior because of the interplay between quantum and thermal fluctuations. This yields a characteristic quantum critical ``fan'' [see Fig.~\ref{schematic}(b)] that has been extensively studied in magnets. This is generically associated with unconventional features, such as non-classical scaling of correlation functions~\cite{lake2005quantum} and breakdown of the quasiparticle picture in metals~\cite{custers2003break}. In addition, quantum critical fluctuations provide a fertile ground for emergence of novel phases, including unconventional superconductivity, for example in heavy fermion metals close to their magnetic quantum critical point~\cite{gegenwart2008quantum}. 

Materials in the vicinity of ferroelectric to paraelectric phase transitions are attracting a renewed interest~\cite{rowley2014ferroelectric}. Quantum criticality theory applied to ferroelectrics~\cite{khmelnit1971low,roussev2003theory} predicts measurable signatures in the dielectric constant, and the quantum paraelectrics SrTiO$_{3}$ and KTaO$_{3}$ indeed exhibit the expected scaling over a broad temperature range~\cite{rowley2014ferroelectric}. Based on these insights, ferroelectric quantum critical behaviour has been suggested as the origin of superconductivity in doped SrTiO$_{3}$, with critical fluctuations of the ferroelectric mode proposed as the ``glue'' for Cooper pairs~\cite{edge2015quantum}. Specifically, it was shown theoretically~\cite{edge2015quantum}, and confirmed experimentally in isotope- and Ca-substituted samples~\cite{stucky2016isotope,rischau2017ferroelectric}, that such a mechanism causes strong enhancement of the superconducting critical temperature as the system is pushed closer to the ferroelectric QCP.

Here we introduce the concept of multiferroic quantum criticality (MFQC), in which magnetic and ferroelectric QCPs occur in the same system. Multiferroic materials, with their coexisting polarization and magnetization, are of fundamental interest due to the coupling between these different orders~\cite{spaldin2005renaissance} and provide an ideal platform for studying the coupling and competition between orders at low temperature. We identify a number of specific material systems within the family of established multiferroics in which MFQC should be realizable and describe the experimental signatures and exotic properties associated with a multiferroic quantum critical point. \\

\noindent \textbf{Phenomenology of multiferroic quantum criticality}

We consider a system described by an effective action of the form~\cite{she2010stability,morice2017hidden}, $S=S_{\phi}+S_{\psi}+S_{\phi\psi}$, where 

\begin{align}
 S_{\phi} = \int d^{d}rd\tau\left[-\alpha_{\phi}|\bm{\phi}|^{2}+\frac{1}{2}\beta_{\phi}|\bm{\phi}|^{4}+\frac{1}{2}|\partial_{\mu}\bm{\phi}|^{2}\right], \label{action1}
\end{align}

\begin{align}
 S_{\psi} = \int d^{d}rd\tau\left[-\alpha_{\psi}|\bm{\psi}|^{2}+\frac{1}{2}\beta_{\psi}|\bm{\psi}|^{4}\right] \nonumber \\ 
 + \int d^{d}kd\omega\left[\frac{k^{2}}{2}+\gamma\frac{\omega}{k^{z-2}}\right]|\bm{\psi}|^{2}. \label{action2}
\end{align}

Here $\bm{\phi}$ and $\bm{\psi}$ are real, multiple-component fields representing the ferroelectric and magnetic order parameters, the index $\mu=0,1,...,d$ runs over time and $d$ spatial dimensions. $r$ and $\tau$ denote position and imaginary time, $k$ and $\omega$ the momentum and frequency, and we choose units such that the propagation speed for both fields is unity. The actions describe continuous ferroelectric and magnetic phase transitions with dynamical exponents equal to one for $\bm{\phi}$ and $z$ for $\bm{\psi}$~\cite{she2010stability,morice2017hidden}. For individual fields, the customary quartic interactions $S_{\phi}^{\mathrm{int}}=\frac{\beta_{\phi}}{2}\int d^{d}rd\tau |\bm{\phi}|^{4}$ and $S_{\psi}^{\mathrm{int}}=\frac{\beta_{\psi}}{2}\int d^{d}rd\tau |\bm{\psi}|^{4}$, give rise to susceptibilities, $\chi$, scaling with temperature as $\chi_{\phi}^{-1}\sim T^{d-1}$ and $\chi_{\psi}^{-1}\sim T^{(d+z-2)/z}$ (for $d+z >4$; small logarithmic corrections are expected at low temperatures for $d+z=4$~\cite{chandra2017prospects}). In three dimensions, this leads to dielectric susceptibility $\chi_{\phi}^{-1}\sim T^{2}$ for a quantum critical ferroelectric and magnetic susceptibility $\chi_{\psi}^{-1}\sim T^{3/2}$ for an antiferromagnet ($z=2$). Note that the dynamics arising from a ferromagnetic QCP ($z=3$) are different from those of an antiferromagnetic one ($z=2$).

When these two QCPs occur in the same system, the separation between them, $\delta$, can be tuned by using an additional control parameter [Fig.~\ref{schematic}(a)], and the interaction between the fields [Fig.~\ref{schematic}(c)-(d)] leads to measurable signatures in scaling of observables as we show below.

First we review the effect of a biquadratic interaction 

\begin{equation}
S_{\phi\psi}^{0}=\frac{g_{0}}{2}\int d^{d}rd\tau |\bm{\phi}|^{2}|\bm{\psi}|^{2},
\label{biquadratic}
\end{equation}

which is allowed by symmetry for all pairs of fields. $S_{\phi\psi}^{0}$ produces additional contributions to the susceptibilities [Fig.~\ref{schematic}(c)-(d)], such that the overall fluctuations at the bicritical point are governed by the phase with the lower exponent~\cite{oliver2015quantum,morice2017hidden}. Therefore, $\chi_{\phi}^{-1}$ shows a crossover from $T^{2}$ to $T^{3/2}$ scaling, as the antiferromagnetic QCP approaches the ferroelectric one [Fig.~\ref{schematic}(g)]. On the other hand, $\chi_{\psi}^{-1}$ continues to follow $T^{3/2}$ scaling, unaffected by the proximity of the ferroelectric QCP. 

We note that such a biquadratic interaction can lead to qualitative changes in the phase diagram. Specifically, the sign of the aforementioned corrections to $\chi_\phi^{-1}$ is determined by $g_0$ itself and therefore can be either positive or negative. Consequently, the proximity of the magnetic QCP can either suppress or enhance the ferroelectric phase and change the scaling of the corresponding $T_c$ as a function of the control parameter as illustrated in Fig.~\ref{schematic}(h). Interestingly, this can lead to a situation in which, by reducing the temperature, the paraelectric phase displays a reentrant behavior due to the MFQC [Fig.~\ref{schematic}(h)].

In our MFQC case, $\bm{\phi}$ is associated with a polar distortion and $\bm{\psi}$ with magnetic order. Consequently, two additional interactions of lower order than the biquadratic are allowed by symmetry. First, a gradient interaction of the form 

\begin{equation}
S_{\phi\psi}^{1}=g_{1}\int d^{d}rd\tau \bm{\phi}\cdot[\bm{\psi}(\nabla\cdot\bm{\psi})-(\bm{\psi}\cdot\nabla)\bm{\psi}]
\end{equation}

corresponding to the inverse Dzyaloshinskii-Moriya interaction responsible for the spin-induced polarization in spiral magnets~\cite{Katsura/Nagaosa/Balatsky:2005,cheong2007multiferroics}. Second, the dynamical counterpart of this interaction, 

\begin{equation}
S_{\phi\psi}^{2}=g_{2}\int d^{d}rd\tau \bm{\psi}\cdot(\bm{\phi} \times \partial_{\tau} \bm{\phi}),
\label{dynamical}
\end{equation}

which can be spin-orbit coupling or exchange driven~\cite{dzyaloshinskii2009intrinsic,juraschek2017dynamical}, becomes important in the vicinity of quantum phase transitions. Using finite-temperature field theory~\cite{abrikosov2012methods}, we find that the $S_{\phi\psi}^{2}$ interaction gives a correction to $\chi_{\phi}^{-1}$ proportional to 

\begin{align}
k_{B}T\sum_{n}\int d^{d}k\frac{\omega_{n}^{2}}{(-\alpha_{\phi}+\frac{k^{2}}{2}+\frac{\omega_{n}^{2}}{2})(-\alpha_{\psi}+\frac{k^{2}}{2}+\gamma\frac{\omega_{n}}{k^{z-2}})} \nonumber \\
=\int d^{d}kk^{z-2}\left[\frac{\omega_{\phi}}{2}\frac{\omega_{\psi}-\omega_{\phi}+2\omega_{\psi}n_{\mathrm{B}}(\omega_{\phi})}{\omega_{\phi}^{2}-\omega_{\psi}^{2}} + \frac{\omega_{\psi}^{2}n_{\mathrm{B}}(\omega_{\psi})}{\omega_{\psi}^{2}-\omega_{\phi}^{2}} \right],
\end{align}

where $\omega_{n}=2\pi nk_{B}T$ ($n=0,\pm 1,...$) are bosonic Matsubara frequencies, $\pm \omega_{\phi}=\pm \sqrt{-\alpha_{\phi}+k^{2}/2}$ and $\omega_{\psi}=\frac{1}{\gamma}(-\alpha_{\psi}k^{z-2}+k^{z}/2)$, with $n_{\mathrm{B}}(\omega)=\frac{1}{e^{\omega/k_{B}T}-1}$ being the Bose function. The corresponding Feynman diagram is shown in Fig.~\ref{schematic}(e). Evaluating the integral over the momenta in three dimensions, we obtain 

\begin{equation}
\chi_{\phi}^{-1}\sim T^{3-1/z},
\label{chi}
\end{equation}

to lowest order in temperature as the two QCPs come close together, remarkably different from the simplest biquadratic coupling case in which the field with the lower exponent dominates. Instead, for proximal ferroelectric and antiferromagnetic ($z=2$) QCPs, we find that these unconventional corrections yield $\chi_{\phi}^{-1}\sim T^{5/2}$ as a subdominant correction to the dielectric susceptibility arising from the influence of magnetic criticality, with the leading term scaling as $T^{2}$. A similar calculation for the magnetic susceptibility, from evaluating the diagram shown in Fig.~\ref{schematic}(f), yields $\chi_{\psi}^{-1}\sim T^{2}$. The spatial gradient term, $S_{\phi\psi}^{1}$, leads to higher order corrections to these scaling expressions ($\chi_{\phi}^{-1}\sim T^{2z}$).

We expect that this interplay between QCPs will lead to similar crossovers in other quantum critical scaling laws. For example, the Gr{\"u}neisen parameter $\Gamma$, which is the ratio of the thermal expansion coefficient to the specific heat~\cite{zhu2003universally}, diverges as $\Gamma\sim T^{-1/\nu z}$ in the quantum critical regime. Therefore multiferroic quantum critical behaviour should manifest as different scaling of $\Gamma$ near antiferromagnetic ($\Gamma\sim T^{-1}$) and ferroelectric ($\Gamma\sim T^{-2}$) QCPs. Note that the exponents of $\Gamma$ obtained from scaling theory or mean field theory are similar at the upper critical dimension~\cite{chandra2017prospects}.

Finally, a special case of ``coupled bicriticality'' occurs when variation of a single control parameter results in coincident magnetic and ferroelectric QCPs, as shown in Fig.~\ref{schematic}(i). To evaluate the behavior in this regime, we use the biquadratic coupled action as the starting point for a renormalization group (RG) analysis treating quantum fluctuations in both fields. Employing an $\epsilon$-expansion and retaining terms to one loop order we obtain the set of flow equations for $\alpha$, $\beta$ and $g_{0}$~\cite{she2010stability}. From an analysis of the flow trajectories, we find that the bicritical point is stable for $g_{0}<\sqrt{\beta_{\phi}\beta_{\psi}}$, whereas larger $g_{0}$ gives runaway flow trajectories, indicative of first-order phase transitions. \\

\noindent \textbf{Material candidates}

\begin{figure}
\includegraphics[scale=0.6]{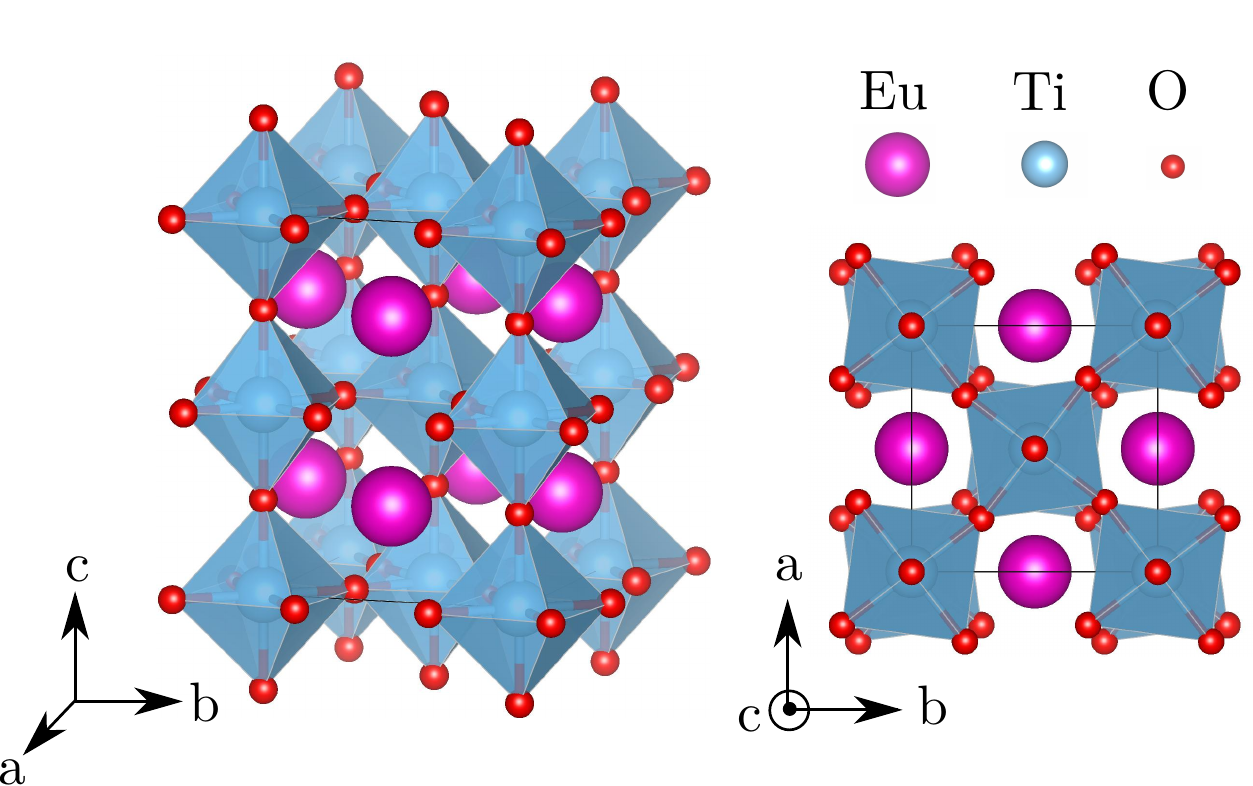}
  \caption{\textbf{Crystal structure of EuTiO$_{3}$.} Europium titanate forms in the perovskite structure with antiferrodistortive rotations around the $c$ axis ($a^0a^0c^-$ in Glazer notation).}  \label{crystal_structure}
\end{figure}

Next, we propose material candidates for achieving MFQC. Europium titanate, EuTiO$_{3}$, is an interesting material that crystallizes in the perovskite structure, shown in Fig.~\ref{crystal_structure}, with antiferrodistortive rotations of the oxygen octahedra leading to a tetragonal $I4/mcm$ space group~\cite{rushchanskii2012first,goian2012antiferrodistortive}. The localized $4f^7$ moments on the Eu$^{2+}$ ions order in a G-type antiferromagnetic arrangement (with all nearest neighbour spins oppositely aligned) at the low N\'{e}el temperature of 5.3 K because of their small inter-ionic exchange interactions~\cite{mcguire1966magnetic}. At the same time, EuTiO$_{3}$ has a large and diverging dielectric constant at low temperatures, indicative of its proximity to a ferroelectric phase transition~\cite{kamba2007magnetodielectric}; this behavior is strikingly similar to that of SrTiO$_{3}$. Recently, Das studied the temperature and magnetic-field dependence of the dielectric susceptibility in EuTiO$_{3}$~\cite{das2012quantum}. Here we extend this work to explore EuTiO$_{3}$ as a model system for MFQC, and describe two strategies, alloying and strain engineering, to engineer multiferroic QCPs.

\begin{figure}
\includegraphics[scale=0.6]{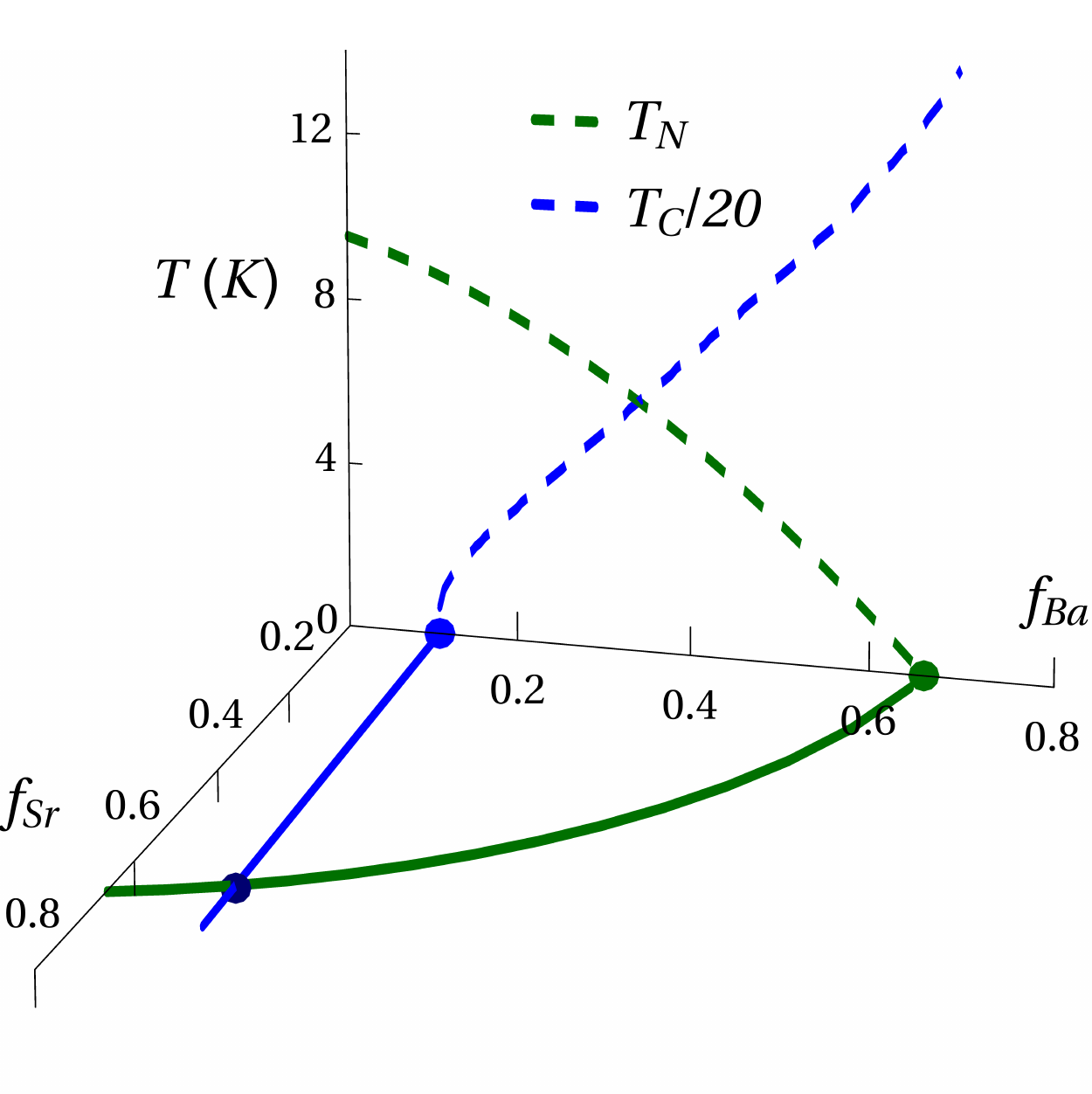}
  \caption{\textbf{Tuning criticality in EuTiO$_{3}$ by alloying.} Temperature-concentration phase diagram of (Eu,Ba,Sr)TiO$_{3}$. Dashed lines show the finite temperature phase boundaries of antiferromagnetic (green) and ferroelectric (blue) phases with increasing Ba concentration, $f_{Ba}$, with no Sr. Solid lines track magnetic (green) and ferroelectric (blue) quantum critical points with increasing Sr concentration, $f_{Sr}$. An ``accidental'' bicritical point is reached at the crossing of the two, at a composition Eu$_{0.3}$Ba$_{0.1}$Sr$_{0.6}$TiO$_{3}$.}  \label{alloying}
\end{figure}

EuTiO$_{3}$ can be readily alloyed with the canonical ferroelectric BaTiO$_{3}$~\cite{rushchanskii2010multiferroic}, which suppresses the magnetic ordering through dilution, at the same time promoting ferroelectricity by increasing the volume and favouring off-centering of ions from their high-symmetry positions~\cite{rushchanskii2010multiferroic}. Alloying with the quantum paraelectric SrTiO$_{3}$~\cite{guguchia2012tuning}, on the other hand, should reduce the N\'{e}el temperature without affecting the ferroelectric properties, since EuTiO$_3$ and SrTiO$_3$ have the same lattice constant. This flexibility allows for a rich multiferroic phase diagram, including control of the position of and separation between the magnetic and ferroelectric QCPs.

To analyze the phase diagram of the (Eu,Ba,Sr)TiO$_3$ system, we performed Ising- and Heisenberg- model simulations with parameters extracted from density functional calculations in this work or measurements reported in the literature (see Methods section for details). The resulting phase diagram is shown in Fig.~\ref{alloying}. First, we examine the effect of Ba concentration, $f_{Ba}$. The N\'{e}el temperature, $T_{N}$, is suppressed to zero giving a magnetic QCP at $f_{Ba}\approx 0.7$, very close to the percolation limit on a cubic lattice. Ferroelectricity emerges from a ferroelectric QCP at $f_{Ba} \approx 0.1$. For intermediate compositions ($0.1<f_{Ba}<0.7$) a multiferroic phase is obtained. Next, we track the positions of the magnetic and ferroelectric QCPs with increasing Sr concentration, $f_{Sr}$, as Sr is substituted for Eu. The ferroelectric QCP remains unchanged with increasing $f_{Sr}$. On the other hand, the magnetic QCP moves to lower values of $f_{Ba}$ as $f_{Sr}$ is increased. Alloying with Sr, therefore tunes the separation between the two QCPs. Interestingly, at $f_{Ba}\approx 0.1$ and $f_{Sr}\approx 0.6$, i.e. for composition Eu$_{0.3}$Ba$_{0.1}$Sr$_{0.6}$TiO$_{3}$, an ``accidental'' bicritical point, at which the magnetic and ferroelectric QCPs are coincident, occurs. 

\begin{figure}
\includegraphics[scale=0.6]{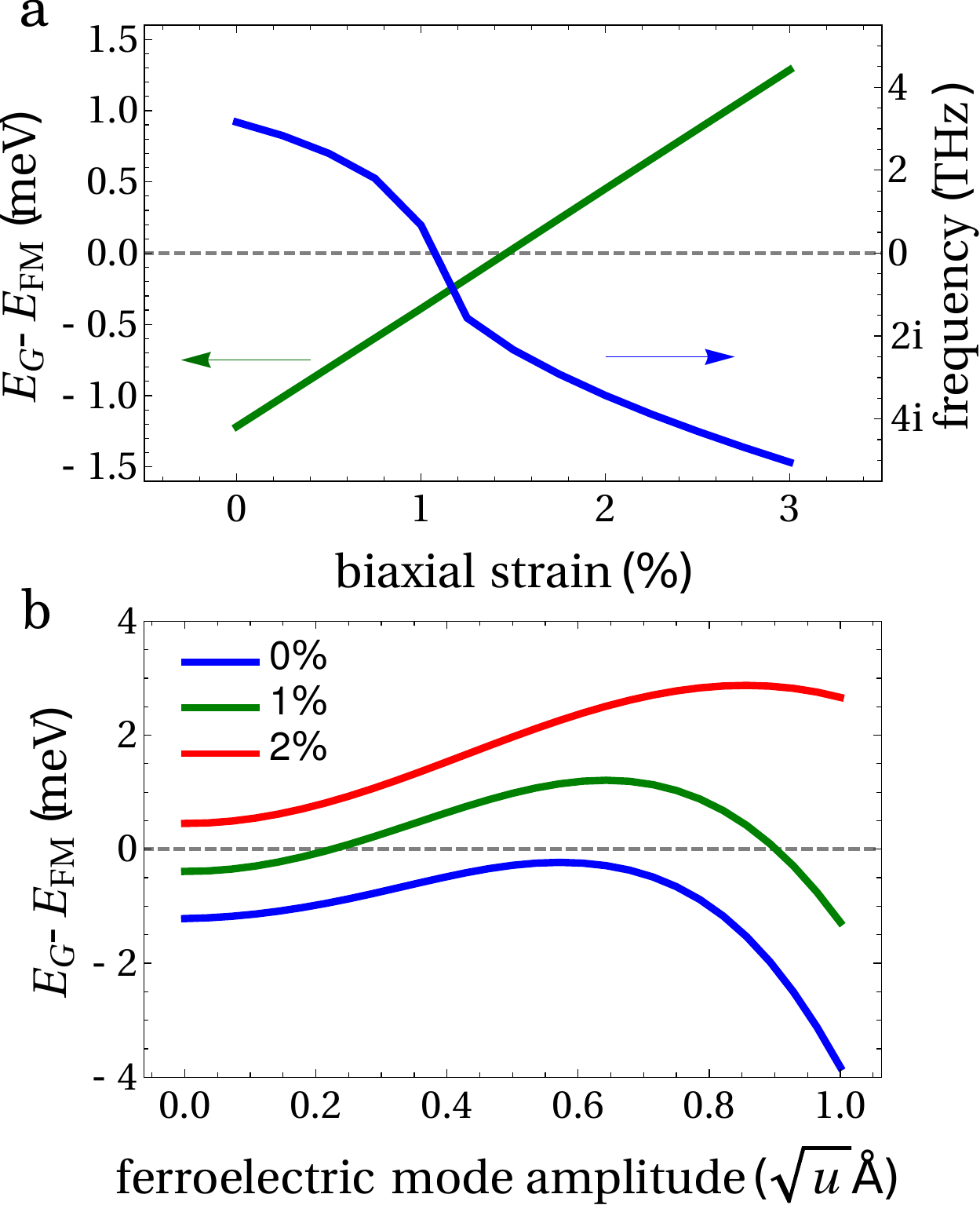}
  \caption{\textbf{Near-bicriticality in strained EuTiO$_{3}$.} (a) Difference in energy between G-type antiferromagnetic and ferromagnetic (FM) orders as a function of biaxial strain (green curve), from density functional calculations. Energy differences are in meV per $I4/mcm$ unit cell. G-type order is favoured for less than $\sim$1.4 \% strain, ferromagnetic order at larger values. Blue curve shows the ferroelectric phonon mode frequency with G-type ordering as a function of strain. Imaginary frequencies correspond to unstable phonon modes, indicative of a transition to a ferroelectric state. (b) Energy difference between G-type and FM orders with increasing amplitude of the ferroelectric mode at different values of strain. Here $u$ is the atomic mass unit. At $\sim$1 \% strain, close to the critical value, the system is both magnetically and structurally ``soft'' and the relative stability of the magnetic orders can be reversed by varying the amplitude of the ferroelectric mode.}  \label{strain}
\end{figure}

\begin{table*}[ht]
\centering
\caption{Candidate materials for multiferroic quantum criticality.}
\label{candidate_materials}
\begin{tabular}{ c c c c }
\hline\hline
Material & Dielectric nature & Magnetic nature & Tuning strategy \\ \hline\hline
EuTiO$_{3}$ & quantum paraelectric & G-type AFM ($T_{N}=5.3$ K) \cite{mcguire1966magnetic} & Ba/Sr alloying at Eu site; strain \\
SrTiO$_{3}$ & quantum paraelectric & diamagnetic \cite{kleemann2009multiglass} & Mn doping at Sr site \\
KTaO$_{3}$ & quantum paraelectric & diamagnetic \cite{kleemann2009multiglass} & Mn doping at K site \\
NaMnF$_{3}$ & quantum paraelectric & G-type AFM ($T_{N}=66$ K) \cite{dubrovin2018unveiling} & Mg/Zn alloying at Mn site \\
TbMnO$_{3}$ & improper ferroelectric & spiral order ($T_{N}=28$ K) \cite{kimura2003magnetic} & Y/Eu alloying at Tb site \\
\hline\hline
\end{tabular}
\end{table*}

As a second possible tuning parameter, we explore strain, which has proved to be a useful tool to tune the properties of perovskite oxide thin films~\cite{schlom2007strain}. Indeed, under biaxial tensile strain imposed via coherent heteroepitaxy, EuTiO$_{3}$ has been reported to become a ferromagnetic ferroelectric~\cite{fennie2006magnetic,lee2010strong}. We performed density functional calculations (see Methods for details) to explore the effect of biaxial strain on the quantum critical behaviour of EuTiO$_{3}$ using the low-temperature tetragonal $I4/mcm$ structure~\cite{rushchanskii2012first,goian2012antiferrodistortive}, in contrast to previous work, which assumed a cubic structure. We found that G-type antiferromagnetic order is favored for unstrained and small values of tensile biaxial strain, and ferromagnetic order becomes stable beyond $\sim$1.4 \% strain [Fig.~\ref{strain}(a)]. Although, the energy differences between different magnetic orders are small, we expect the general trends to be accurately described by density functional calculations.  At a similar, though (in contrast to previous calculations~\cite{fennie2006magnetic}) not identical, value of critical strain, the ferroelectric phonon mode frequency becomes imaginary, indicating a ferroelectric instability. While biaxially strained EuTiO$_{3}$ is not precisely bicritical, the onset of ferroelectricity promotes ferromagnetic order, and our RG analysis indicates that a first-order quantum phase transition will occur. The strong influence of the fluctuations in one order parameter on the other in this nearly bicritical scenario is strikingly revealed in Fig.~\ref{strain}(b), where we plot the energy difference between the G-type antiferromagnetic and ferromagnetic orders with increasing ferroelectric mode amplitude. The magnetic energy difference is very sensitive to the amplitude of the ferroelectric mode, especially close to the critical strain, where the relative stability of G-type antiferromagnetic and ferromagnetic order is reversed with increasing ferroelectric mode amplitude.

In Table~\ref{candidate_materials} we suggest other candidate materials for exploring multiferroic quantum criticality. Mn doping at the Sr or K sites in quantum paraelectric SrTiO$_{3}$ and KTaO$_{3}$ has been reported to result in a ``magnetoelectric multiglass'' state with simultaneous polar and magnetic glassy states~\cite{kleemann2009multiglass}. With appropriate choice of doping, these systems could be tuned to MFQC. Going beyond oxides, perovskite fluoride NaMnF$_{3}$ has been recently found to show quantum paraelectricity coexisting with G-type antiferromagnetic order~\cite{dubrovin2018unveiling}. Suppressing the N\'{e}el temperature by Mg or Zn alloying at the Mn site could tune the magnetic sublattice to quantum criticality. Improper ferroelectrics, in which polarization arises as a secondary effect of a magnetic (or other lattice) ordering, are a promising class of materials to search for quantum bicritical multiferroics~\cite{kimura2003magnetic}. For example, in the canonical improper ferroelectric TbMnO$_{3}$, increasing the size of the rare earth ion by alloying with Eu changes the magnetic ground state from a spiral order to a collinear one, which drives a simultaneous quantum phase transition from a ferroelectric to a paraelectric state~\cite{ishiwata2010perovskite,fedorova2018relationship}. The dynamical coupling term, $S_{\phi\psi}^{2}$, is likely to dominate in this case, leading to our predicted unconventional scaling laws.

A promising route to engineer new multiferroics has been through heterostructures and nanostructures~\cite{ramesh2007multiferroics}; such atomic-scale engineering could be used to create quantum critical multiferroics. For instance, the N\'{e}el temperature of EuTiO$_{3}$/SrTiO$_{3}$ superlattices could be tuned with the relative number of layers of the two components and reach criticality. Heterostructures of materials with magnetic QCPs (such as heavy fermions) and materials with ferroelectric QCPs (for example SrTiO$_{3}$) could be used to achieve controllable and tunable composite or interfacial MFQC. With reduced spatial dimensions such systems could prove to be interesting playgrounds for exploring novel scaling laws below the upper critical dimension, distinct from bulk materials. \\

\noindent \textbf{Implications and prospects}

The effects of the coupled magnetoelectric critical excitations proposed here should manifest in an experimentally accessible temperature range. The temperature scales for observing universal quantum critical properties are related to the N\'{e}el temperature ($\sim$ 5K in the case of EuTiO$_3$) for the magnetic part and the Debye temperature for the ferroelectric case. For the closely related quantum paraelectrics, SrTiO$_{3}$ and KTaO$_{3}$, the experimentally reported limit is $\sim$50 K~\cite{rowley2014ferroelectric}. 

Adding charge carriers would provide an additional non-thermal control parameter. From our preliminary calculations we find, for example, that doping of $5\times 10^{19}$ electrons/cm$^{3}$ moves 1\% biaxially strained EuTiO$_{3}$ to a magnetic QCP while hardening the ferroelectric phonon mode due to enhanced electrostatic screening between the electric dipoles. Similar trends of stabilization of ferromagnetic order and suppression of ferroelectricity with electron doping, are also found in alloyed EuTiO$_{3}$. The presence of carriers could also lead to emergent modulated order within which QCPs are hidden, as has been observed recently in NbFe$_{2}$~\cite{friedemann2018quantum}. More excitingly, carriers offer the intriguing possibility of Fermi surface instabilities emerging around the critical points, potentially enabling superconductivity. Close to an antiferromagnetic QCP, $d$-wave superconductivity is expected on a cubic lattice~\cite{miyake1986spin,scalapino1986d}. In contrast, a ferroelectric QCP would give rise to $s$-wave pairing~\cite{edge2015quantum}. Such an interplay between different kinds of pairing, arising from distinct QCPs, is an interesting avenue for future exploration. 

Since ferroelectric and magnetic phase transitions can be discontinuous, it is important to determine the nature of the phase transitions in alloyed and strained EuTiO$_{3}$ and our other proposed systems. For example, perovskite SrTiO$_{3}$ and KTaO$_{3}$ show continuous phase transitions and the resulting quantum criticality~\cite{rowley2014ferroelectric}, whereas a critical end point was recently observed at a magnetic-field-induced metaelectric phase transition in multiferroic BiMn$_{2}$O$_{5}$~\cite{kim2009observation}. While weakly first-order quantum phase transitions still lead to nearly critical fluctuations driving quantum criticality and could allow for observation of our proposals, first-order multiferroic quantum phase transitions could additionally be interesting in their own right.  In this context, a recently proposed mechanism of ``quantum annealed criticality''~\cite{chandra2018quantum}, in which first-order finite-temperature phase transitions can end in a zero temperature critical point, could perhaps also be explored in our multiferroic quantum critical scenario. Discontinuous magnetic quantum phase transitions often result in exotic phases such as magnetic rotons, instantons and skyrmion textures~\cite{pfleiderer2005first}, and the implications associated with an additional breaking of space-inversion symmetry remain to be explored.

Criticality with multiple order parameters can also be engineered in ultra-cold quantum gases. An exciting development in this direction is the recent demonstration of coupling and competition between two order parameters in a Bose-Einstein condensate coupled to two optical cavities~\cite{morales2017coupling}. We envisage that our predicted critical scaling crossovers arising from coupled order parameters of different types could also be explored in next-generation quantum gas-optical cavity systems.

In summary, we introduced the concept of multiferroic quantum criticality, which combines magnetic and ferroelectric quantum critical behaviour in the same system. We described the phenomenology of multiferroic quantum criticality, discussed its implications and presented suitable systems and schemes to realize it. Our work is particularly timely given the recent surge of interest in quantum materials~\cite{basov2017towards}, and we hope that our findings motivate the exploration of coupling and competition between various quantum critical behaviors.

\section*{Methods}

\noindent \textbf{Finite-temperature field theory calculations:} We write the free propagators for the two fields  as

\begin{equation}
 G_{\phi}=\frac{1}{-\alpha_{\phi}+k^{2}/2+\omega^{2}/2}
\end{equation}

\begin{equation}
 G_{\psi}=\frac{1}{-\alpha_{\psi}+k^{2}/2+\gamma\omega/k^{z-2}}
\end{equation}

The $S_{\phi\psi}^{2}$ interaction gives a correction to $\chi_{\phi}^{-1}$ of the form

\begin{align}
& k_{B}T\sum_{n}\int d^{d}k G_{\phi}\omega_{n}^{2}G_{\psi} \nonumber \\
& = k_{B}T\sum_{n}\int d^{d}k\frac{\omega_{n}^{2}}{(-\alpha_{\phi}+\frac{k^{2}}{2}+\frac{\omega_{n}^{2}}{2})(-\alpha_{\psi}+\frac{k^{2}}{2}+\gamma\frac{\omega_{n}}{k^{z-2}})}
\end{align}

We carry out the summation over the Matsubara frequencies using the standard contour integral technique~\cite{abrikosov2012methods}, yielding 

\begin{equation}
 \int d^{d}kk^{z-2}\left[\frac{\omega_{\phi}(\omega_{\psi}-\omega_{\phi})}{2(\omega_{\phi}^{2}-\omega_{\psi}^{2})}+\frac{\omega_{\phi}\omega_{\psi}n_{\mathrm{B}}(\omega_{\phi})}{\omega_{\phi}^{2}-\omega_{\psi}^{2}} + \frac{\omega_{\psi}^{2}n_{\mathrm{B}}(\omega_{\psi})}{\omega_{\psi}^{2}-\omega_{\phi}^{2}} \right].
\end{equation}

Close to criticality ($\alpha_{\phi},\alpha_{\psi}\rightarrow 0$) in three dimensions, the third term gives the lowest exponent in the temperature dependence of $T^{3-1/z}$. This leads to the strongest correction to $\chi_{\phi}^{-1}$, as presented in equation~\ref{chi} of the main text. 

\noindent \textbf{Density functional calculations:} Our first-principles calculations were carried out using the Vienna Ab-initio Simulation Package ({\sc vasp})~\cite{kresse1996efficient}, with the Perdew-Burke-Ernzerhof approximation to the exchange correlation functional~\cite{perdew1996generalized}. Eu $4f$ electrons were treated with the GGA+$U$ method, using Dudarev's approach~\cite{dudarev1998electron}, with $U=6.0$ eV and $J=1.0$ eV. Default projector augmented wave pseudopotentials were employed. A plane wave cutoff of 500 eV was used and the Brillouin zone was sampled using an $8\times 8\times 6$ $k$-point grid. Phonon calculations were performed using the {\sc phonopy} code~\cite{togo2015first}, with 80 atom supercells using a $4\times 4\times 6$ $k$-point mesh. For the biaxially strained EuTiO$_{3}$, the strain tensor reads

\begin{equation}
 \varepsilon=\begin{pmatrix}
              \zeta & 0 & 0 \\
              0 & \zeta & 0 \\
              0 & 0 & -\nu\zeta \\
             \end{pmatrix},
\end{equation}

where $\zeta=(a-a_{0})/a_{0}$ is the applied strain ($a_{0}$ and $a$ are the equilibrium and strained in-plane lattice constants) and $\nu$ is the biaxial Poisson ratio.

\noindent \textbf{Ising model for estimating ferroelectric critical temperature:} Ferroelectric alloys were modeled by a simple transverse Ising model~\cite{blinc1974soft}, which has been shown to give reasonable estimates for experimental critical temperatures~\cite{zhang2000study},

\begin{align}
 H = -\Omega\sum_{i}\sigma_{i}^{x}-\frac{1}{2}\sum_{ij}J_{ij}\sigma_{i}^{z}\sigma_{j}^{z} \nonumber \\
 -\frac{1}{4} \sum_{ijkl}J_{ijkl}\sigma_{i}^{z}\sigma_{j}^{z}\sigma_{k}^{z}\sigma_{l}^{z}-2\mu E\sum_{i}\sigma_{i}^{z}.
\end{align}

Here $\Omega$ is the tunneling frequency and $\mu$ is the effective dipole moment, which couples to the external electric field $E$. $\sigma_{i}^{x,y,z}$ are pseudospins at the $i$-th site, which interact via two-body ($J_{ij}$) and four-body ($J_{ijkl}$) exchange terms. The term proportional to $\sigma_{x}$ is the tunneling between the two minima of the free energy double well, and does not imply that the electric dipole has any precessional dynamics. Alloying is simulated by weighting the parameters by concentration of the constituents $f_{\alpha}$,

\begin{equation}
 J = \sum_{\alpha}f_{\alpha}J_{\alpha}, \quad \Omega = \sum_{\alpha}f_{\alpha}\Omega_{\alpha}, \quad
 \mu = \sum_{\alpha}f_{\alpha}\mu_{\alpha}.
\end{equation}

The polarization is then given by $P=2n\mu\sum_{i}\langle \sigma_{i}^{z}\rangle$, where $n$ is the number of dipoles per unit volume. The change of lattice parameter, $a$, with alloying is approximated using Vegard's law. Treating the pseudospin in the mean field approximation yields a self-consistent equation for the polarization, which was then solved numerically. The susceptibility $\chi=\left(\partial \langle P\rangle/\partial E\right)_{E=0}$ was used to estimate the critical temperatures for different alloy compositions. Parameters used (shown in Table~\ref{ising_parameters}) were previously fitted to reproduce experimental values of critical temperatures and give good estimates of the experimental critical temperatures~\cite{zhang2000study}. 

\begin{table}[h]
\centering
\caption{Parameters used for estimating ferroelectric critical temperature of Ba and Sr alloyed EuTiO$_{3}$.}
\label{ising_parameters}
\begin{tabular}{ c c c c c c}
\hline\hline
Compound & $J_{ij}$(meV) & $J_{ijkl}$(meV) & $\Omega$(meV) & $\mu$(e\AA{})& $a$(\AA{})\\
\hline\hline
BaTiO$_{3}$ & 23.90 & 62.16 & 30.58 & 2.17 & 4.005 \\
SrTiO$_{3}$/EuTiO$_{3}$ & 2.04 & 0 & 6.86 & 1.51 & 3.905\\
\hline\hline
\end{tabular}
\end{table}

\noindent \textbf{Heisenberg model for estimating magnetic critical temperature:} The energies obtained from density functional calculations were mapped to a classical Heisenberg model to calculate the exchange parameters. The $4f^{7}$ moments on the Eu$^{2+}$ are well localized, therefore the system can be reasonably described by a simple Heisenberg model. Alloying of non-magnetic ions (Sr and Ba) was modeled by introducing random binary variables $\zeta_{i}$ for each site $i$, such that

\begin{eqnarray}
 \zeta_{i} &=& 1 \quad i=\mathrm{Eu} \nonumber \\
 &=& 0 \quad i=\mathrm{Sr,Ba}.
\end{eqnarray}

This yields the following Hamiltonian

\begin{equation}
 H=-\sum_{ij}\mathcal{J}_{ij}\zeta_{i}\zeta_{j}S_{i}\cdot S_{j} + \sum_{i}\mathcal{D}_{i}\zeta_{i}S_{i}^{2},
\end{equation}

where $S_{i}$ are classical spins at site $i$, $\mathcal{J}_{ij}$ is the nearest-neighbour exchange interaction strength and $\mathcal{D}_{i}$ is the single ion anisotropy energy. A competition between the exchange term, anisotropy term and dilution through alloying leads to a phase transition by tuning the $\alpha_{\psi}$ coefficient in the action (equation~\ref{action2}). From density functional calculations, exchange interaction strengths were obtained to be: $\mathcal{J}^{ab}$=-0.0286 meV ($ab$-plane exchange parameter) and $\mathcal{J}^{c}$=0.0331 meV ($c$ direction exchange parameter). The magnetic phases and critical temperatures of the system were then estimated using a standard Metropolis-based Monte Carlo procedure~\cite{landau2014guide}. This simple treatment of disorder has previously been successfully applied to dilute magnetic semiconductors~\cite{bergqvist2004magnetic}.

\section*{Acknowledgements}

We acknowledge helpful discussions with G. Aeppli, T. Donner, K. Dunnett, C. Ederer, A. Edstr\"{o}m, T. Esslinger, N. Fedorova, C. Gattinoni, Q. Meier, A. Morales, R. Pisarev and P. Zupancic. This work is supported by ETH-Zurich (AN, AC and NAS) and the US DOE BES E3B7, the Villum foundation, and Knut and Alice Wallenberg Foundation (AVB). Calculations were performed at the Swiss National Supercomputing Centre (project ID p504).

\section*{Author Contributions}

NAS conceived the concept. NAS, AVB, AC and AN devised the analysis. AN carried out the calculations. AN and NAS wrote the manuscript with contributions from all authors.

\bibliography{references}

\end{document}